\documentclass[a4paper]{jpconf}
\usepackage{graphicx}
\usepackage{float}
\usepackage{lineno}
\begin{document}
\title{Probing the QCD phase diagram with the measurements of $\phi$-meson
  production and elliptic flow in the heavy-ion collision at STAR}

\author{Md. Nasim (for the STAR Collaboration)}

\address{National Institute of Science Education and Research (NISER),
\\ Bhubaneswar-751005, India.}
\ead{nasim@rcf.rhic.bnl.gov/mnasim2008@gmail.com}

\begin{abstract}
 We present the measurements of the $\phi$-meson production and elliptic flow ($v_{2}$) at mid-rapidity in Au + Au
  collisions at $\sqrt{s_{NN}}$ = 7.7 - 200 GeV. The data are collected using the STAR detector in the years 2010 and 2011. The energy dependence of nuclear
  modification factor ($R_{\rm{CP}}$) of $\phi$ meson is
  presented. The $\phi$-meson $R_{\rm{CP}}$ has a value $\geq$ 1.0 for
  $\sqrt{s_{NN}}$ $\leq$ 39 GeV. The  $\Omega/\phi$ ratios are also
  presented and show a different
  trend at the intermediate transverse momentum ($p_{T}$) for
  $\sqrt{s_{NN}}$ = 11.5 GeV  compared to higher beam energies. The number-of-constituent quark (NCQ) scaling of $v_{2}$ has been studied at various
beam energies. The NCQ scaling holds for particles and anti-particles
separately including the $\phi$ meson for
$\sqrt{s_{NN}}$ $\geq$ 19.6 GeV, which can be considered as an evidence of
partonic collectivity.  We observe 
at $\sqrt{s_{NN}}$ = 7.7 and 11.5 GeV,  the $\phi$-meson $v_{2}$ falls off the
trend from the other hadrons at highest measured $p_{ T}$ 
values by 1.8$\sigma$ and 2.3$\sigma$, respectively.

\end{abstract}

\section{Introduction}
The $\phi$ vector meson is the lightest bound state of a strange (s) quark
and a anti strange ($\bar{s}$) quark. It has a mass of 1.01945 $\pm$
0.00002 GeV/$c^{2}$ which is comparable to the mass of lightest baryons
 $p$ (0.938 GeV/$c^{2}$) and $\Lambda$ (1.115 GeV/$c^{2}$)~\cite{pdg}. The life time of the $\phi$ meson is $\sim$ 42
fm/$c$. Because of longer life time, the $\phi$ meson will mostly decay outside
the fireball and therefore its daughters will not have much time to
re-scatter in the hadronic phase. The hadronic interaction cross-section
of the $\phi$ meson is expected to have a small value and hence seems to freeze out early compared to other lighter hadrons ($\pi$, $K$ and $p$)~\cite{white}. 
Therefore its production should be less affected by the later
stage hadronic interactions in the evolution of the system formed in
heavy-ion collisions.\\
The elliptic flow parameter $v_{2}$ is a good tool for studying
the system formed in the early stages of high energy collisions
at RHIC~\cite{hydro}. It describes the momentum anisotropy of particle
emission from non-central heavy-ion collisions. It is defined as the
second harmonic coefficient of the Fourier decomposition of azimuthal distribution  with respect to the reaction plane angle
($\Psi$) and can be written as
\begin{equation}
v_{2}=\langle\cos(2(\phi-\Psi))\rangle,
\end{equation}
where $\phi$ is emission azimuthal angle~\cite{method}. Although the
elliptic flow is early time phenomena but its magnitude might still be
affected by the later stage hadronic interaction. Since the hadronic
 interaction cross section of $\phi$ meson is smaller than the other
 hadrons, its $v_{2}$ remains almost unaffected by the late stage
 interaction~\cite{BN,NBN}. Therefore $\phi$-meson $v_{2}$ will reflect the collective
 motion of the partonic phase. These make the $\phi$ meson  a clean
probe for the study of the QCD phase diagram in the Beam Energy Scan (BES)
program at the Relativistic Heavy Ion Collider (RHIC)~\cite{starnote}.

\section{Data Sets and Analysis Details}
The results presented here are based on data collected from Au+Au collisions at
$\sqrt{s_{NN}}$=  7.7, 11.5, 19.6, 27, 39, 62.4 and 200 GeV with the
STAR detector for minimum bias trigger in the years of 2010 and 2011. The cuts on primary vertex position along the longitudinal
beam direction ($V_{z}$) is 30 cm for 200 GeV, 40 cm for 39 and 62.4 GeV, 50 cm for 11.5
GeV and 70 cm for 7.7, 19.6 and 27 GeV GeV data set. An additional cut
on vertex radius $<$ 2 cm  has been used
to reject contamination from beam pipe interaction. The Time
Projection Chamber (TPC)
and Time of Flight (TOF) detectors 
with full azimuthal coverage were used for particle identification in the
central pseudo-rapidity ($\eta$) region ($|\eta|<$ 1.0). For spectra
analysis we have used only the TPC  where as for elliptic flow analysis both
TPC and TOF detectors have been used. $\phi$-mesons are identified using the invariant mass technique from
their decay  to $K^{+} + K^{-}$ (branching ratio is 49.04 $\pm$ 0.6
$\%$). Mixed event technique has been used for combinatorial
background estimation~\cite{phi_plb_star}. The $\eta$-sub event plane
method with a $\eta$ gap of $|\Delta\eta|$ $<$ 0.05 has been used for elliptic flow measurements~\cite{method}.

\section{Results}

\subsection{ \bf{Transverse momentum spectra}}
We present $R_{\rm{CP}}(0-10\%/40-60\%)$ measurement of $\phi$-meson at mid-rapidity
($|\it y|$ $<$ 0.5) in Au+Au collisions at $\sqrt{s_{NN}}$ = 7.7 - 39
GeV.
The $R_{\rm{CP}}$ is defined as the ratio of the particles yield in the central to
peripheral collisions normalized by number of inelastic binary
collisions ($N_{bin}$). The value of  $N_{bin}$ is calculated from the
Monte Carlo Glauber simulation~\cite{Nbin}. If nucleus-nucleus collisions are simply a superposition of nucleon-nucleon
collisions then $R_{\rm{CP}}$ is equal to one. Deviation of $R_{\rm{CP}}$ from the unity would imply
contribution fron the nuclear medium effects. Because of the
energy loss of the partons traversing the high density QCD medium, the
$R_{\rm{CP}}$ of $\phi$ mesons goes below unity at 200 GeV~\cite{white}. From Fig. 1, one can
see at the intermediate $p_{T}$ ($p_{T}$ $>$ 1.5 GeV/c), $R_{\rm{CP}}$
$\geq$ 1.0 for $\sqrt{s_{NN}}$ $\leq$ 39 GeV. This indicates that at low energy the parton
energy loss effect is less important.
\begin{figure}[!ht]
\begin{center}
\includegraphics[scale=0.28]{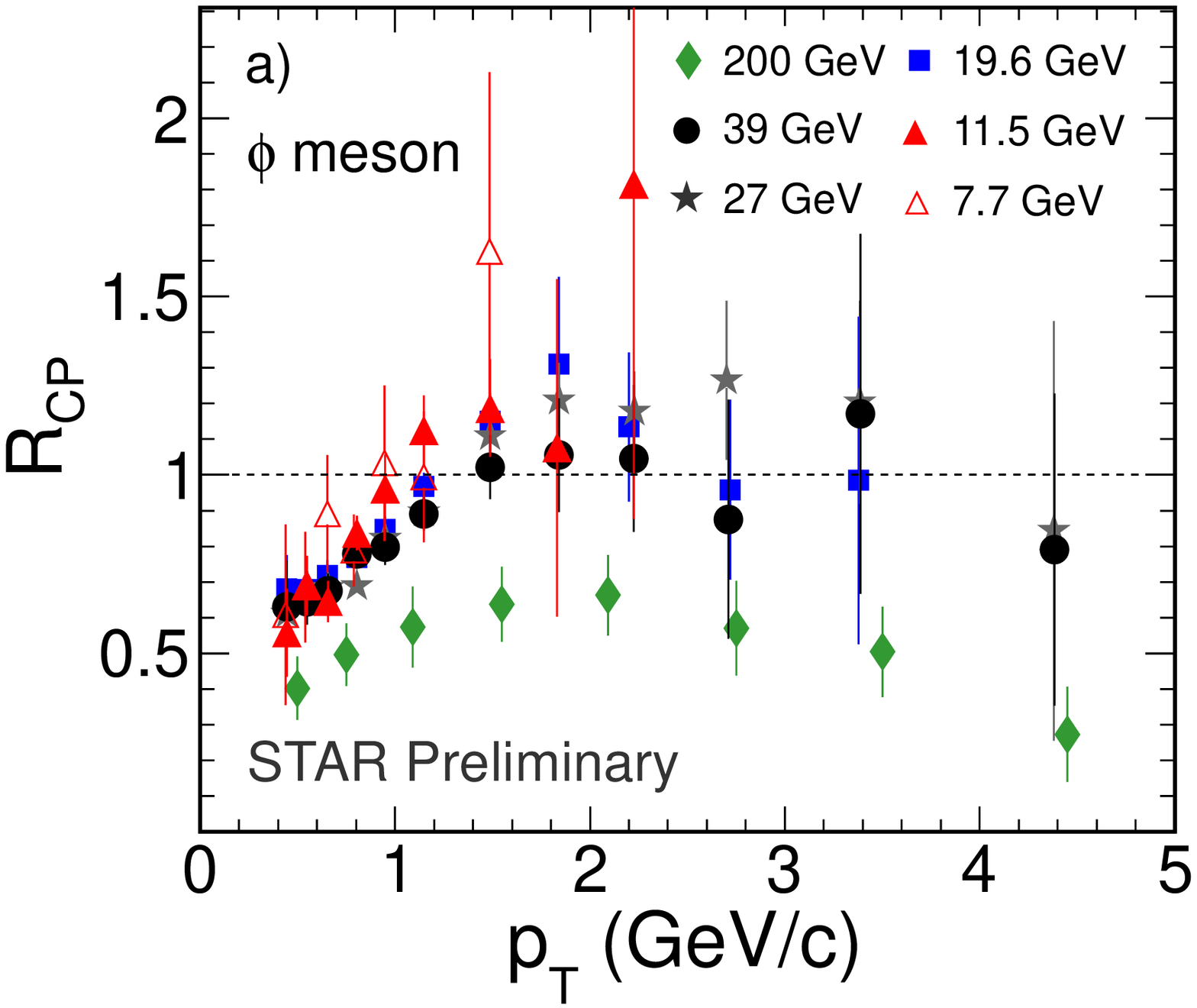}
\includegraphics[scale=0.235]{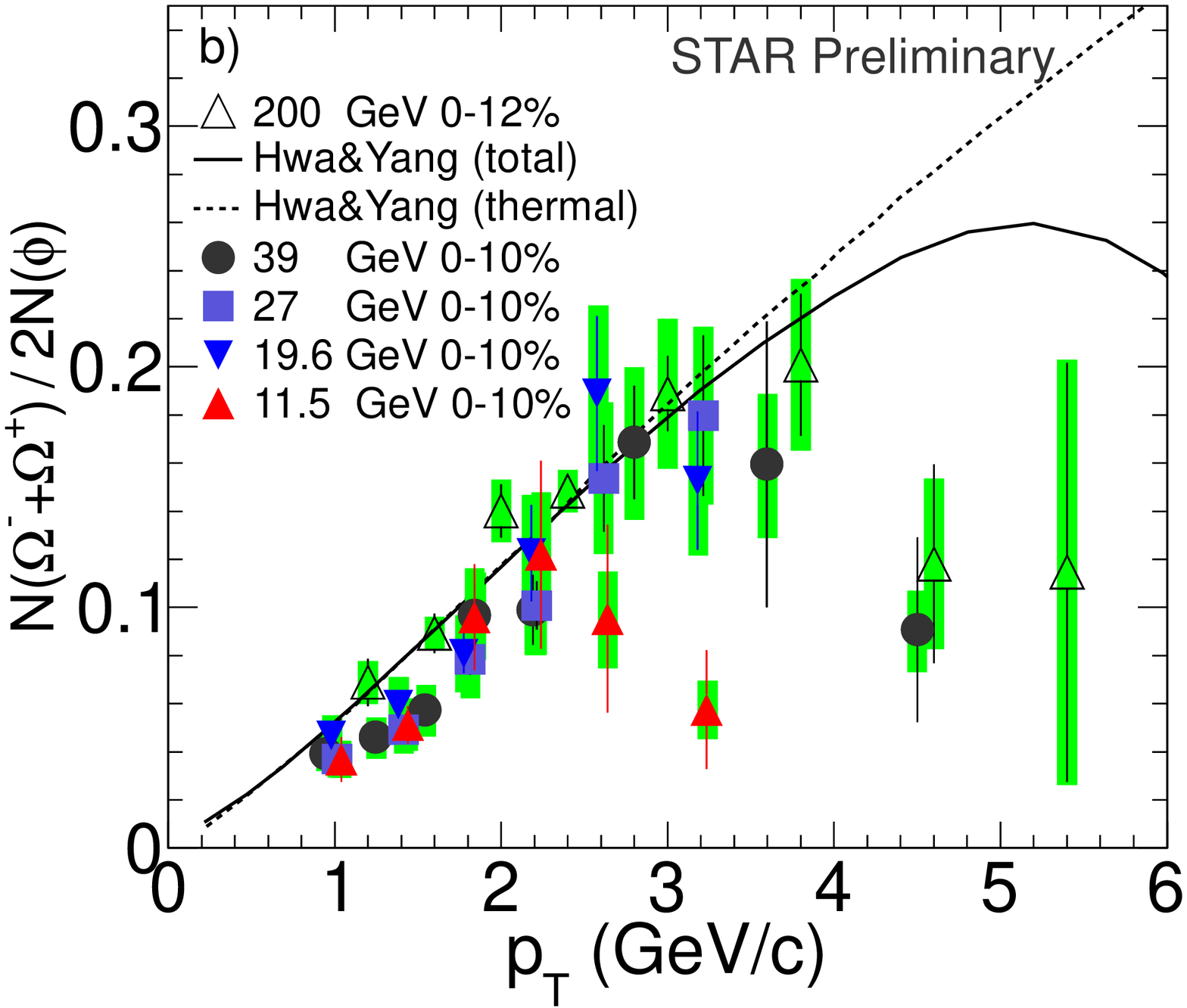}
\caption{(Color online) Left panel : $R_{\rm{CP}}$ as function of
  $p_{T}$ in the Au+Au collision
  at various beam energies.  The $R_{\rm{CP}}(0-05\%/40-60\%)$ at
  $\sqrt{s_{NN}}$ = 200 GeV are taken from previous STAR measurements~\cite{phi_star_prc}. Error bars are only
  statistical uncertainties. Right panel : ($\Omega^{-} +
  \overline{\Omega}^{+}$)/2$\phi$ as a function of $p_{T}$.  Green
  bands are the systematic error and vertical lines are statistical
  error.}
\label{fig1}
\end{center}
\end{figure} \\

The panel (b) of Fig. 1 shows the baryon-to-meson ratio, N($\Omega^{-}
+ \overline{\Omega}^{+}$)/2N($\phi$), as a function of $p_{T}$ in Au + Au collisions
at $\sqrt{s_{NN}}$ = 11.5 GeV to 200 GeV. The data points for 200 GeV
are taken from Ref.~\cite{phi_star_prc}. The dashed lines are the
results from the recombination model calculations with thermal strange quarks~\cite{thermal_quark}. 
In Au+Au central collisions at $\sqrt{s_{NN}}$ = 200 GeV, the ratios
of N($\Omega^{-} + \overline{\Omega}^{+}$)/2N($\phi$) in the intermediate $p_{T}$ range are explained by the
recombination model with thermal strange quarks. The ratios N($\Omega^{-} + \overline{\Omega}^{+}$)/2N($\phi$) for 
$\sqrt{s_{NN}}$ $\geq$ 19.6 GeV show similar trend. But at
$\sqrt{s_{NN}}$ = 11.5 GeV, the ratio at the highest measured $p_{T}$
shows a deviation from the trend of other energies. The $\chi^{2}$/ndf for
deviation between 11.5 and 19.6 GeV is $\sim$ 8.3/2 for $p_{T}$ $>$
2.4 GeV/c. This may suggest a change in $\Omega$ and/or $\phi$ production
mechanism at $\sqrt{s_{NN}}$ = 11.5 GeV.\\

\subsection{\bf{Elliptic flow}}
\begin{figure}[!ht]
\begin{center}
\includegraphics[scale=0.28]{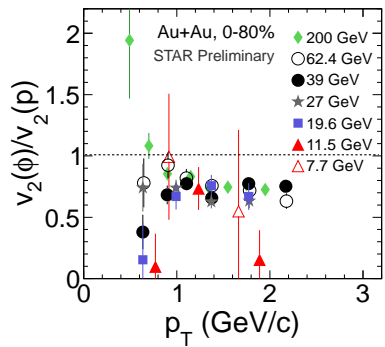}
\caption{(Color online) $v_{2}(\phi)/v_{2}(p)$ ratio as function of
  $p_{T}$ in the Au+Au collision
  at various beam energies for 0-80$\%$ centrality~\cite{prc,prl}. Error bars are only
  statistical uncertainties.}
\label{fig2}
\end{center}
\end{figure} 

Figure 2 shows $v_{2}(\phi)/v_{2}(p)$ ratios as function of $p_{T}$ for
all beam energies from $\sqrt{s_{NN}}$ = 7.7 GeV to 200 GeV. We
observe that at low $p_{T}$  the $v_{2}(\phi)/v_{2}(p)$ ratio
decreases with decrease in beam energy. This could be due to
smaller partonic collectivity at low energy, since the $\phi$
$v_{2}$ mostly reflect collectivity from the partonic phase. On the
other hand if we look at  200 GeV, the situation is different. One can
see that at top RHIC energy $v_{2}(\phi)/v_{2}(p)$ $>$ 1.0 at
low $p_{T}$ ($p_{T}$ $<$ 0.7 GeV/c). This can be explained due to later stage hadronic
interaction effect on proton $v_{2}$ as predicted in the theoretical (hydro + hadron cascade)
model~\cite{hyrdo_cascade}.

\begin{figure}[!ht]
\begin{center}
\centerline{\includegraphics[scale=0.55]{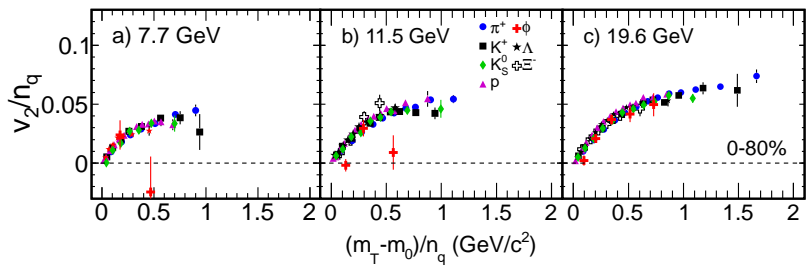}}
\caption{(Color online) The elliptic flow ($v_{2}$) scaled by
  number-of-constituent quark ($n_{q}$) as a
  function of $(m_{T}-m_{0})/n_{q}$ for selected particles in the Au+Au collision
  at various beam energies for 0-80$\%$ centrality~\cite{prc,prl}. Error bars are only
  statistical uncertainties.}
\label{fig5}
\end{center}
\end{figure} 
Figure 3 shows $v_{2}$ divided by number-of-constituent quark ($n_{\rm{q}}$)  as
function  of  $(m_{T} - m_{0})/n_{\rm{q}}$, where $m_{T} ( =\sqrt{p_{T}^{2}
+ m_{0}^{2}})$ is the transverse mass and $m_{0}$ is the mass of the hadron, at $\sqrt{s_{NN}}$ =
7.7, 11.5 and 19.6 GeV~\cite{prc,prl}. The NCQ
scaling holds fairly well for $\sqrt{s_{NN}}$ $\geq$ 19.6 GeV for particles and 
anti-particles separately including the $\phi$ meson (results for $\sqrt{s_{NN}}$ $>$ 19.6 GeV are
not shown here). This could
be considered as a signature of partonic collectivity~\cite{ncq1} . However, we observe
at $\sqrt{s_{NN}}$ = 7.7 and 11.5 GeV  the $\phi$-meson $v_{2}$ falls off the
trend from the other hadrons at highest measured  $p_{ T}$
values by 1.8$\sigma$ and 2.3$\sigma$, respectively. Due to the small hadronic interaction cross-section,
 $v_{2}$ of $\phi$ meson mostly reflect collectivity from the partonic
phase~\cite{BN,NBN}. So the small magnitude of the $\phi$-meson $v_{2}$ at
$\sqrt{s_{NN}}$ $\leq$ 11.5 GeV could be the effect of a matter, where
hadronic interactions are more important but we need more statistics to
make strong conclusion. 

\section{Summary}
We report the study of $\phi$-meson production and elliptic flow at mid-rapidity in Au + Au
collisions at $\sqrt{s_{NN}}$ = 7.7 - 200 GeV recorded by the STAR
detector. At the intermediate $p_{T}$, the nuclear modification factor $R_{\rm{CP}}$ of $\phi$ increases with decreasing beam energies, indicating that the partonic
energy loss effect becomes less important at lower beam
energies. The ratios of ($\Omega^{-} + \overline{\Omega}^{+}$)/2$\phi$ in the intermediate
$p_{T}$ range show a different trend at 11.5 GeV compared to those for
the higher beam energies. This may suggest a change of particle production mechanism at lower beam energy. The NCQ scaling holds for
$\sqrt{s_{NN}}$ $\geq$ 19.6 GeV for particle and  anti-particle separately. We observe at $\sqrt{s_{NN}}$ $\leq$11.5 GeV  the $\phi$-meson $v_{2}$
show deviation ($\sim$ 2$\sigma$) from the other hadrons at highest measured  $p_{ T}$
values and also the $v_{2}(\phi)/v_{2}(p)$ ratios at low $p_{T}$
decreases with decrease in beam energy. This may indicate
that the contribution to the collectivity from 
partonic phases decreases at lower beam energies. 
\section{Acknowledgements}
Financial support from DST International Travel Support and DST SwarnaJayanti project, Government of India is gratefully acknowledged.

\end{document}